\documentclass[journal]{IEEEtran}
\usepackage{mathrsfs}
\usepackage{bbm}
\usepackage{amssymb}
\usepackage{bbding}
\usepackage{threeparttable}

\usepackage[mathcal]{euscript}

\usepackage{psfrag,calc,url,bm}

\usepackage{cite}

\usepackage{graphicx}

\usepackage{psfrag}

\usepackage{subfigure}

\usepackage{url}

\usepackage{stfloats}

\usepackage{amsmath}

\usepackage{float}

\usepackage{algorithmic}

\usepackage[ruled,linesnumbered,vlined]{algorithm2e}

\usepackage{color}

\usepackage{boxedminipage}

\usepackage{amsthm}

\usepackage{multirow}

\usepackage{setspace}

\usepackage{soul}
\usepackage{epstopdf}

\allowdisplaybreaks[3]

\usepackage{multirow}
\usepackage{amsmath}
\usepackage{booktabs}

\usepackage{makecell}
\usepackage[table]{xcolor}

\begin{document}

\title{RIS-Assisted Downlink Pinching-Antenna Systems: GNN-Enabled Optimization Approaches}
\author{Changpeng He, Yang Lu,~\IEEEmembership{Member,~IEEE}, Yanqing Xu,~\IEEEmembership{Member,~IEEE}, Chong-Yung Chi,~\IEEEmembership{Fellow,~IEEE}, \\and~Arumugam Nallanathan,~\IEEEmembership{Fellow,~IEEE}, 
\thanks{Changpeng He and Yang Lu are with the State Key Laboratory of Advanced Rail Autonomous Operation, and also with the School of Computer Science and Technology, Beijing Jiaotong University, Beijing 100044, China (e-mail: 25110135@bjtu.edu.cn, yanglu@bjtu.edu.cn).}
\thanks{Yanqing Xu is with the School of Science and Engineering, The Chinese University of Hong Kong, Shenzhen, 518172, China (email: xuyanqing@cuhk.edu.cn).}
\thanks{Chong-Yung Chi is with the Institute of Communications Engineering, Department of Electrical Engineering, National Tsing Hua University, Hsinchu 30013, Taiwan (e-mail: cychi@ee.nthu.edu.tw).}
\thanks{Arumugam Nallanathan is with the School of Electronic Engineering and Computer Science, Queen Mary University of London, London and also with the Department of Electronic Engineering, Kyung Hee University, Yongin-si, Gyeonggi-do 17104, South Korea (e-mail: a.nallanathan@qmul.ac.uk).}
}

\maketitle

\begin{abstract}
This paper investigates a reconfigurable intelligent surface (RIS)-assisted multi-waveguide pinching-antenna (PA) system (PASS) for multi-user downlink information transmission, motivated by the unknown impact of the integration of emerging PASS and RIS on wireless communications. First, we formulate sum rate (SR) and energy efficiency (EE) maximization  problems in a unified framework, subject to constraints on the movable region of PAs, total power budget, and tunable phase of RIS elements. Then, by leveraging a  graph-structured topology of the RIS-assisted PASS, a novel three-stage graph neural network (GNN) is proposed, which learns PA positions based on user locations, and RIS phase shifts according to composite channel conditions at the first two stages, respectively, and finally determines beamforming vectors. Specifically, the proposed GNN is achieved through unsupervised training, together with three implementation strategies for its integration with convex optimization, thus offering trade-offs between inference time and solution optimality. Extensive numerical results are provided to validate the effectiveness of the proposed GNN, and to support its unique attributes of viable generalization capability, good performance reliability, and real-time applicability. Moreover, the impact of key parameters on RIS-assisted PASS is illustrated and analyzed.

\end{abstract}


\section{Introduction}

The evolution toward sixth-generation (6G) wireless networks demands unprecedented data rates, ultra-low latency, and exceptional energy efficiency (EE) to support emerging applications such as holographic communications, digital twins, and the tactile internet \cite{wang2023road}. To meet these stringent requirements, novel programmable metasurfaces, which can intelligently reconfigure the wireless propagation environment, have emerged as an essential technology. Among these metasurfaces, the reconfigurable intelligent surfaces (RIS) \cite{di2020smart1, wu2021intelligent} and the pinching-antenna (PA) systems (PASS) \cite{ding2024pinching,Liu2024pinching} stand out as two promising candidates, offering complementary advantages for future wireless networks. On one hand, RIS utilizes a large array of passive reflecting elements with adjustable phase shifts to intelligently construct cascade wireless links to enable  signal strength enhancement, interference suppression, coverage extension, and  obstacle bypassing \cite{basar2019wireless}. The passive nature of RIS elements ensures low power consumption and cost-effective deployment, making it particularly suitable for energy-constrained scenarios \cite{liu2021reconfigurable}. Recent studies have demonstrated significant performance gains in RIS-assisted communication systems, including improved spectral efficiency and enhanced physical-layer security\cite{huang2019reconfigurable, guan2020intelligent}. On the other hand, PASS introduces additional spatial degrees of freedom via the flexible mobility of PAs. Unlike conventional fixed-position antenna arrays, PAs can dynamically adjust their locations along their waveguides to exploit favorable channel conditions. Particularly, PAs can be placed in close proximity to users to efficiently reduce path loss. Recent research has shown that PASS can achieve further substantial performance improvements, especially in scenarios with severe channel fading or limited spatial diversity \cite{ding2024pinching}.



Given that RIS and PASS are deployable independently of each other, one natural approach to unlocking synergistic integration gains is to integrate these two technologies into a unified system \cite{RISvsPA}. Nevertheless, this integration introduces formidable optimization challenges, as the resulting joint optimization problem involves strongly coupled high-dimensional decision variables, including PA positions, RIS phase shifts, and transmit beamforming vectors. Traditional optimization approaches, such as alternating optimization (AO) and successive convex approximation (SCA), typically decompose the joint optimization problem into multiple subproblems and solve them in a round-robin manner  \cite{yu2020robust, pan2021intelligent}. This decomposition often fails to preserve the joint optimization gains fully.  Furthermore, these methods suffer from prohibitive computational complexity, which scales poorly with increasing system scales, rendering them impractical for time-varying wireless networks\cite{cvx-iterations}.

Deep learning (DL) has emerged as a transformative paradigm for tackling complex optimization problems in wireless communications \cite{zhang2019deep, sun2022learning}. By training neural networks to learn the desired mapping from system parameters to near-optimal solutions, DL-based approaches can achieve an orders-of-magnitude speedup in inference time compared to iterative algorithms. Early studies leveraged multilayer perceptrons (MLPs) to optimize beamforming vectors and resource allocation in various wireless scenarios. For instance, MLP-based approaches have been proposed for movable antenna array systems to jointly learn antenna positions and beamforming vectors \cite{Related Work MLP}. However, MLPs may fail to capture the key features underlying network topologies and inter-user interactions. This limitation results in poor generalization capability when the network configuration changes, and requires retraining to adapt to different problem sizes \cite{shen2020graph}. Graph neural networks (GNNs) address these limitations by explicitly representing wireless networks as graphs, where nodes correspond to network entities (e.g., base stations, users, RIS elements) and edges capture inter-node relationships (e.g., communication links, interference) \cite{GNN for gr}. Through message passing mechanisms, GNNs can effectively extract structural features and learn complex inter-node dependencies while maintaining scalability across varying network sizes \cite{shen2023gnn}. Recent studies have demonstrated the superiority of GNNs over traditional neural networks in various wireless optimization tasks, including power control \cite{shen2020graph}, spectrum sharing \cite{zhang2025optimal}, and beamforming design \cite{li2023deep}. 


Such an emerging RIS-assisted PASS, which may significantly upgrade the transmission capacity, has yet to be justified. To this end, we propose a novel three-stage GNN in a unified framework to address both sum-rate (SR) and EE maximization for its downlink transmission. To the best of our knowledge, the proposed GNN is the first effective solution. The major contributions are summarized as follows:
\begin{itemize}
\item We formulate SR and EE maximization problems for a RIS-assisted multi-waveguide PASS by jointly optimizing PA positions, RIS phase shifts, and transmit beamforming vectors, under constraints of minimum PA spacing along waveguides, unit-modulus RIS phase shifts, and total power budget.

\item The proposed three-stage GNN facilitates joint optimization of the considered system. The three-stage GNN  decomposes the joint problem into three sequential but interdependent stages. By explicitly modeling the wireless system as a graph, the GNN effectively leverages users' positions to configure PA positions and RIS phase shifts, and captures inter-user interactions to determine transmission strategies. It also incorporates customized activation functions and constraint-satisfaction mechanisms that guarantee the feasibility of learned solutions.

\item In view of the lack of optimality guarantees for the three-stage GNN, we integrate the GNN with conventional optimization methods. Specifically, three implementation strategies, including a fully learning-based approach, and two hybrid learning-optimization approaches, thereby exhibiting trade-offs between solution effectiveness and inference efficiency among them.

\item Numerical results validate the effectiveness of jointly leveraging RIS and PA in enhancing system SR and EE. The three-stage GNN’s fast inference speed and viable generalization capability with respect to unseen problem sizes are justified. By comparing the three strategies, we validate the effectiveness of each stage in optimizing PA positions, RIS phase shifts, and beamforming vectors. Moreover, the impact of key system parameters on system performance is illustrated and discussed.


\end{itemize}
\emph{Notation}: The following mathematical notations and symbols are used throughout this paper. $\bf a$ and $\bf A$ stand for a column vector and a matrix (including multiple-dimensional tensor), respectively. The sets of real numbers, and $n$-by-$m$ real matrices are denoted by ${\mathbb{R}}$, and ${\mathbb{R}^{n \times m}}$, respectively. The sets of $n$-dimensional complex column vector, and $n$-by-$m$ complex matrices are denoted by ${\mathbb{C}^n}$, and ${\mathbb{C}^{n \times m}}$, respectively. For a complex number $a$, $\left| a \right|$ denotes its modulus and ${{\rm Re}(a)}$ denotes its real part. For a vector $\bf a$, ${\left\| \bf a \right\|}$ is the Euclidean norm. For a matrix ${\bf A}$, ${\bf A}^H$ and $  \left \|{\bf A}\right\|$ denote its conjugate transpose and Frobenius norm, respectively. $[{\bf a}]_i$, $[{\bf A}]_{i,j}$, and $[{\bf A}]_{i,:}$  are the $i$-th element of the vector ${\bf a}$, the $i$-th row and the $j$-th column element of the matrix $\bf A$, and the $i$-th row vector of the matrix $\bf A$, respectively. $\{a_{i}\}$ denotes the set of elements  for all the admissible $i$. $\mathcal{CN}(\cdot)$ denotes the circularly symmetric complex Gaussian distribution.

\section{Related Works}

 This section reviews relevant research across three main areas: RIS-assisted communication systems, PASS, and GNN-enabled wireless designs.

\subsection{RIS-Assisted Communication Systems}

The optimization of RIS-assisted wireless systems has attracted substantial research attention due to the promise of intelligent environment control. Early works focused on jointly optimizing transmit beamforming vectors and RIS phase shifts within AO frameworks. For instance, Huang et al. \cite{yu2020robust} proposed a robust AO algorithm for jointly designing active beamforming vectors, artificial noise, and RIS phase shifts to enhance physical-layer security under imperfect eavesdropper channel state information (CSI). Huang et al. \cite{huang2019reconfigurable} investigated energy-efficient downlink multi-user communication aided by RIS, proposing AO algorithms that jointly optimize transmit power allocation and RIS phase shifts using gradient descent and fractional programming. While these convex optimization-based methods can guarantee convergence to stationary points, their computational complexity scales poorly with the number of RIS elements and users, making real-time implementation challenging \cite{pan2021intelligent}. To reduce computational complexity, researchers have explored DL approaches for RIS optimization. Deep reinforcement learning-based methods \cite{huang2021deep} have been proposed to infer RIS phase shifts from CSI, achieving a significant speedup compared to iterative algorithms. Jiang et al. \cite{jiang2021learning} developed a GNN-based approach that jointly learns beamforming vectors and RIS phase shifts with implicit channel estimation capability, demonstrating robustness to channel estimation errors. However, these works consider transmitters equipped  with fixed-position antennas.

\subsection{Pinching-Antenna Systems} 

 Recent research has explored various optimization techniques for PASS across different communication scenarios and performance objectives. Lv et al. \cite{lv2025beam} proposed a three-stage beam training scheme for the near-field PASS, where a coarse location was first obtained with one activated PA, phase matching was then achieved with an increased number of PAs, and precise beam alignment was finally realized through exhaustive search. Hou et al. \cite{hou2025performance} investigated uplink PASS performance and optimized PA positions to maximize channel gains. Zhu et al. \cite{zhu2025secure} addressed physical-layer security in PASS-enabled communications by developing a PA-wise successive tuning algorithm that ensures constructive signal superposition at legitimate users while inducing destructive signals at eavesdroppers, and proposed two transmission architectures with artificial noise for multi-waveguide scenarios. While these optimization-based and algorithmic methods provide convergence guarantees, they suffer from high iterative computational complexity.


Furthermore, DL-based approaches for PA optimization have been recently proposed. Xie et al. \cite{xie2025graph} developed a bipartite graph attention network (BGAT) that formulates the downlink PASS as a bipartite graph and jointly optimizes PA placement and power allocation for EE maximization through unsupervised learning. However, this work only considered single-waveguide scenarios. For multi-waveguide configurations, Kang et al. \cite{kang2025campass} proposed CaMPASS-Net, a dual-stream DL framework with residual connections, for capacity maximization of PASS through the joint optimization precoding and PA positions. Guo et al. \cite{guo2025gpass} introduced GPASS, where one sub-GNN first learns PA positions and another subsequently optimizes transmit beamforming vectors. Despite these advances in learning-based PASS optimization, existing works are not extensible to  considered RIS-assisted PASS. The synergistic combination of PAs and RIS could potentially yield significant performance gains by simultaneously adapting both the transmitter structure and the wireless channel characteristics.

\subsection{GNN-Enabled Wireless Designs} 

GNNs have emerged as a powerful tool for wireless resource allocation due to their ability to capture the network topology and inter-user relationships \cite{lu2025agentic}. Shen et al. \cite{shen2020graph} pioneered the application of GNNs to wireless power control, demonstrating that GNN-based policies can achieve near-optimal performance while generalizing to different network sizes. Eisen and Ribeiro \cite{eisen2020optimal} extended this work to large-scale wireless resource allocation using random edge GNNs, proving that GNN can provide theoretical optimality guarantees under certain conditions. These foundational works established the viability of GNNs for wireless optimization problems.

For beamforming designs, several GNN-based approaches have been proposed. Li et al. \cite{li2023deep} developed a graph attention network (GAT) for energy-efficient beamforming in multi-user downlink systems, incorporating complex-valued operations to directly process channel coefficients. The same authors extended their work to SR maximization \cite{li2024gnn} and max-min fairness problems \cite{li2024gnnmaxmin}, demonstrating the versatility of GNN architectures across different objectives. Chen et al. \cite{chen2021gnn} proposed an unsupervised GNN-based beamforming scheme that decouples power allocation from beamforming vector computation, reducing the output dimensionality and improving learning efficiency. Song et al. \cite{song2024deep} developed GNN frameworks for physical-layer secure beamforming in multi-antenna systems.

Recent works have also explored multi-stage GNN architectures for complex wireless optimization problems with deeply coupled variables. Wang et al. \cite{wang2024gnn} proposed a two-stage GNN for joint unmanned aerial vehicle (UAV) placement and transmission design, where the first stage determined UAV locations and the second stage optimized beamforming vectors. He et al. \cite{he2025fasnet} proposed a two-stage GNN for fluid antenna systems, where the first stage learned antenna positions and the second stage optimized beamforming vectors through unsupervised learning, demonstrating that the two stages can function both jointly and separately. These multi-stage approaches align well with the decomposition structure of joint optimization problems and have shown superior performance compared to single-stage GNN architectures.


\section{System Model and Problem Definition}

\begin{figure}[t]
{\centering
{\includegraphics[ width=.48\textwidth]{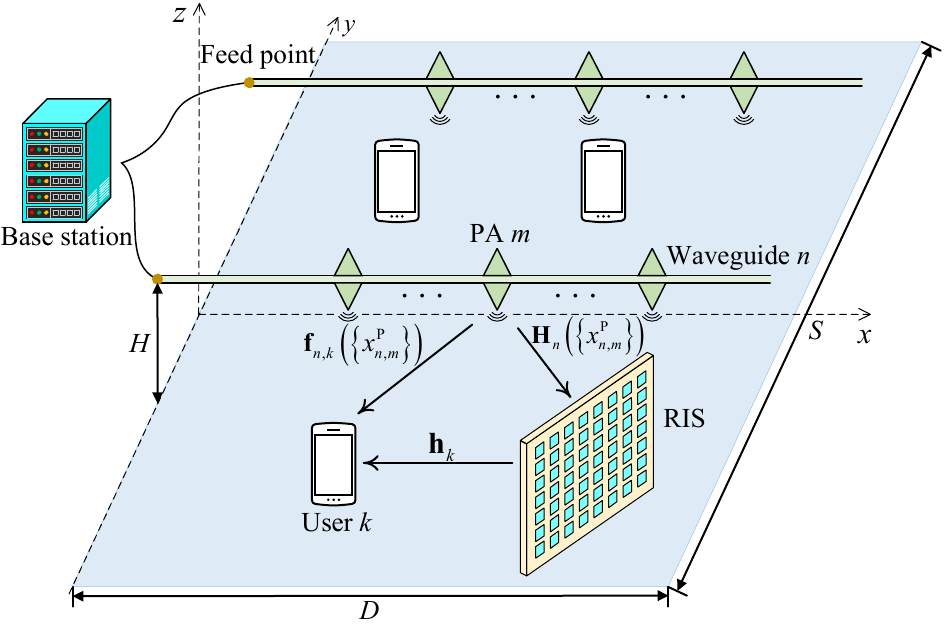}}}
\caption{Illustration of RIS-assisted PASS for downlink multi-user communications.} 
\label{sys}
\end{figure}

We consider a downlink RIS-assisted PASS system within a rectangle region, which spans a length of $D$ along the $x$-axis and $S$ along the $y$-axis. As illustrated in Figure \ref{sys}, a base station, equipped  with $N$ parallel waveguides, serves $K$ ($K\le N$) distinct single-antenna users with the assistance of a RIS. The waveguides are positioned at a height of $H$, extend along the $x$-axis, and span an identical distance of $D$. Each waveguide hosts $M$ movable PAs. We denote the position of PA $m$ on waveguide $n$ as $\bm{\psi}_{n,m}^{\rm P}=(x_{n,m}^{\rm P},y_n^{\rm P},H)$. User $k$ is located at  $\bm{\psi}_k^{\rm U}=(x_k^{\rm U},y_k^{\rm U},0)$ within the $S\times D$ rectangular region. The RIS, comprising $L$ reflecting elements, is located at $\bm{\psi}^{\rm R} = (x^{\rm R},\,y^{\rm R},\,z^{\rm R})$. 

The transmitted signal of each PA is a phase-shifted replica  of the signal from the feed point of its waveguide. The  signal emitted by PAs intended for user $k$ is given by
\begin{flalign}
    \mathbf{s}_k = \mathbf{G}\left(\left\{{x_{n,m}^{\rm P}}\right\}\right) \mathbf{w}_k s_k\in\mathbb C^{MN},
\end{flalign}
where $s_k\in \mathbb{C}$ represents the information symbol, $\mathbf{w}_k \in \mathbb{C}^{N}$ denotes the baseband beamforming vector, and $\mathbf{G}(\{{x_{n,m}^{\rm P}}\})\in{\mathbb C}^{\left(M N\right) \times N}$ denotes the pinching beamforming matrix:
\begin{equation}\label{eqg1}
    \mathbf{G}\left(\left\{{x_{n,m}^{\rm P}}\right\}\right) = \begin{bmatrix}
        \mathbf{g}_1 \left(\left\{{x_{1,m}^{\rm P}}\right\}\right)& \cdots & 0 \\
        \vdots & \ddots & \vdots \\
        0 & \cdots & \mathbf{g}_N\left(\left\{{x_{N,m}^{\rm P}}\right\}\right)
    \end{bmatrix},
\end{equation}
where 
\begin{flalign}\label{eqg2}
    &\mathbf{g}_n\left(\left\{x^{\rm P}_{n,m}\right\}\right) = \\
    &\left[e^{-j\tfrac{2\pi}{\lambda_g}\|\bm{\psi}_{n,0}^{\rm P} - \bm{\psi}_{n,1}^{\rm P}\|}, \dots, e^{-j\tfrac{2\pi}{\lambda_g}\|\bm{\psi}_{n,0}^{\rm P} - \bm{\psi}_{n,M}^{\rm P}\|}\right]^T\in{\mathbb C}^M,\nonumber
\end{flalign}
where $\bm{\psi}^{\rm P}_{n,0} = (0, y^{\rm P}_n, H)$ denotes the location of the feed point for waveguide $n$, and $\lambda_{\rm g}=\lambda/n_{\rm neff}$ denotes the guided wavelength. Here, $\lambda$ is the free-space wavelength corresponding to the carrier frequency, and $n_{\rm neff}$ is the effective refractive index of the dielectric waveguide.

\subsection{Channel Model}

The considered system comprises three kinds of links{\footnote{The proposed three-stage model is extensible to other channel models.}}: 1) the PA-user link, 2) the PA-RIS link, and 3) the RIS-user link, which are detailed as follows.

\subsubsection{PA-user link}

The PA-user link for user $k$ can be modeled by
\begin{flalign}
    &\mathbf{f}_k\left(\left\{{x_{n,m}^{\rm P}}\right\}\right) = \label{direct channel1}\\
    &\left[\mathbf{f}_{1,k}^T\left(\left\{{x_{1,m}^{\rm P}}\right\}\right), \ldots, \mathbf{f}_{N,k}^T\left(\left\{{x_{N,m}^{\rm P}}\right\}\right)\right]^T \in {\mathbb C}^{MN}\nonumber
\end{flalign}
where 
\begin{flalign}\label{direct channel2}
&\mathbf{f}_{n,k}\left(\left\{{x_{n,m}^{\rm P}}\right\}\right) =\\
    &\left[\frac{\sqrt{\eta} e^{-j\tfrac{2\pi}{\lambda}\|\bm{\psi}_k^{\rm U} - \bm{\psi}_{n,1}^{\rm P}\|}}{\|\bm{\psi}_k^{\rm U} - \bm{\psi}_{n,1}^{\rm P}\|}, \dots, \frac{\sqrt{\eta} e^{-j\tfrac{2\pi}{\lambda}\|\bm{\psi}_k^{\rm U} - \bm{\psi}_{n,M}^{\rm P}\|}}{\|\bm{\psi}_k^{\rm U} - \bm{\psi}_{n,M}^{\rm P}\|}\right]^T\in {\mathbb C}^{M},\nonumber
\end{flalign}
denotes the channel vector from $M$ PAs on waveguide $n$ to user $k$ 
where $\eta = c^2/ (4 \pi f_c)^2$ with $c$ being the speed of light and $f_c$ being the free-space carrier frequency.


\subsubsection{PA-RIS Link}

\begin{figure*}[t]
\begin{flalign}\label{ris channel1}
{\bf H}_n\left(\left\{{x_{n,m}^{\rm P}}\right\}\right)= \left[\sqrt{\frac{\beta_0}{{\|\bm{\psi}^{\rm R} - \bm{\psi}_{n,1}^{\rm P}\|^{\alpha}}}}{\bf l}\left({x_{n,1}^{\rm P}}\right),\ldots, \sqrt{\frac{\beta_0}{{\|\bm{\psi}^{\rm R} - \bm{\psi}_{n,M}^{\rm P}\|^{\alpha}}}}{\bf l}\left({x_{n,M}^{\rm P}}\right) \right]\in{\mathbb C}^{L\times M}
\end{flalign}
\hrule
\end{figure*}

The channel between $M$ PAs on  waveguide $n$ and the RIS is modeled by \eqref{ris channel1}, where $\alpha$ denotes the fading exponent, $\beta_0$ denotes the channel gain at the reference distance of $1$ m, and ${\bf l}({x_{n,m}^{\rm P}})\in{\mathbb C}^{L}$ denotes the small fading component from PA $m$ on waveguide $n$ to $L$ reflecting elements, i.e., 
\begin{flalign}\label{ris channel2}
    {\bf l}\left({x_{n,m}^{\rm P}}\right)=\sqrt{\frac{\kappa}{1+\kappa}}{\bf l}^{\rm LoS}\left({x_{n,m}^{\rm P}}\right)+\sqrt{\frac{1}{1+\kappa}}{\bf l}^{\rm NLoS},
\end{flalign}
where $\kappa$ denotes the Rician factor, and
\begin{flalign}\label{ris channel3}
    {\bf l}^{\rm LoS}\left({x_{n,m}^{\rm P}}\right) = \left[1,e^{-j\frac{2\pi}{\lambda}\Delta\varphi_{n,m}},\cdots,e^{-j\frac{2\pi}{\lambda}\left(L-1\right)\Delta\varphi_{n,m}} \right]^T,
\end{flalign}
where $\varphi_{n,m}\in[0,2\pi)$ denotes the cosine of angle-of-departure (AoD) between PA $m$ on waveguide $n$ to the RIS\footnote{We assume that the angle of departure (AoD) from one PA to all reflecting elements of the RIS is identical. This is because the size of the RIS is typically negligible compared to the transmission distance.}, $\Delta$ denotes the element separation, and ${\bf l}^{\rm NLoS}$ contains the non-line-of-sight (NLoS) coefficients, which is circularly symmetric complex Gaussian distributed with zero mean and unity variance. 

Then, the channel matrix from all PAs to the RIS can be expressed as 
\begin{flalign}\label{ris channel4}
{\bf H}\left(\left\{{x_{n,m}^{\rm P}}\right\}\right) = {\rm Concat}\left(\left\{{\bf H}_n\left(\left\{{x_{n,m}^{\rm P}}\right\}\right)\right\}\right)\in{\mathbb C}^{L\times MN},
\end{flalign}
where ${\rm Concat}(\cdot)$ denotes the concatenation operation.


\subsubsection{RIS-user Link}
The channel between the RIS and user $k$ is denoted by $\mathbf{h}_k{\in\mathbb C}^L$, which is given by  
\begin{flalign}
    \mathbf{h}_k = \sqrt{\frac{\beta_0}{{\|\bm{\psi}^{\rm R} - \bm{\psi}_{k}^{\rm U}\|^{\alpha}}}}\left( \sqrt{\frac{\kappa}{1+\kappa}}{\bf h}_{k}^{\rm LoS}+\sqrt{\frac{1}{1+\kappa}}{\bf h}_{k}^{\rm NLoS}\right),
\end{flalign} 
where ${\bf h}_{k}^{\rm LoS}$ and ${\bf h}_{k}^{\rm NLoS}$ denote the LoS and NLoS small fading components, respectively, with a similar definition to those of ${\bf l}^{\rm LoS}({x_{n,m}^{\rm P}})$ and ${\bf l}^{\rm NLoS}$.

\subsection{Problem Formulation}

The received signal at user $k$ via the RIS is given by
\begin{flalign}\label{r_pris}
&\widetilde{y}_{k}= \underbrace {\left( {{\bf f}_k^H\left(\left\{{x_{n,m}^{\rm P}}\right\}\right)+{\bf{h}}_{k}^H{\bm\Phi} {\bf H}\left(\left\{{x_{n,m}^{\rm P}}\right\}\right)} \right)}_{\triangleq \widetilde{\bf h}_k^H\left(\left\{x^{\rm P}_{n,m}\right\}, {\bm\Phi}\right)}\sum\nolimits_{{k^{\prime}}=1}^K {\bf s}_{k^{\prime}} + {n}_{k},
\end{flalign}
where the diagonal matrix
\begin{flalign}\label{rismatrix}
{\bm\Phi} \triangleq {\rm diag}\left\{\left[e^{j\phi_{1}},e^{j\phi_{2}},...,e^{j\phi_{L}}\right]\right\}\in {\mathbb C}^{L\times L} 
\end{flalign}
denotes the phase-shift matrix where $\phi_{l} \in [0,2\pi) $ is the reflecting coefficient of element $l$, and $n_{k}\sim\mathcal{CN}( {0,{\sigma_{k} ^2}})$ denote the additive white Gaussian noises (AWGN). With \eqref{r_pris}, the received information rate at user $k$ is given by \eqref{r_u_p}.

\begin{figure*}[t]    \begin{flalign}\label{r_u_p}
{R}_k\left(\left\{x^{\rm P}_{n,m}\right\}, {\left\{ {{{\bf{w}}_i}} \right\},{\bm\Phi}} \right) ={\log _2}\left( {1 + \frac{{{{\left| {{\widetilde{\bf h}^H_k\left(\left\{x^{\rm P}_{n,m}\right\},{\bm\Phi}\right)}{\bf G}\left(\left\{{x_{n,m}^{\rm P}}\right\}\right){{\bf{w}}_k}} \right|}^2}}}{{\sum\nolimits_{k^{\prime} = 1,{k^{\prime} \ne k}}^K {{{\left| {{\widetilde{\bf h}^H_k\left(\left\{x^{\rm P}_{n,m}\right\},{\bm\Phi}\right)}{\bf G}\left(\left\{{x_{n,m}^{\rm P}}\right\}\right){{\bf{w}}_{k^{\prime}}}} \right|}^2}}+ \sigma _k^2}}} \right)
\end{flalign}
\hrule
\end{figure*}

Our goal is to optimize baseband beamforming vectors, PA positions, and reflecting coefficients to maximize the SR or EE of the considered system, which can be formulated as 
\begin{subequations}\label{p1}
\begin{align}
    {\rm P1}: &\max_{\left\{x^{\rm P}_{n,m}\right\}, {\left\{ {{{\bf{w}}_i}} \right\},{\bm\Phi}}}   {\sum\nolimits_{k=1}^K { R}_k\left(\left\{x^{\rm P}_{n,m}\right\}, {\left\{ {{{\bf{w}}_i}} \right\},{\bm\Phi}} \right)} \label{p1:2a1}\\
    {\rm P2}: &\max_{\left\{x^{\rm P}_{n,m}\right\}, {\left\{ {{{\bf{w}}_i}} \right\},{\bm\Phi}}}   \frac{\sum_{k=1}^K { R}_k\left(\left\{x^{\rm P}_{n,m}\right\}, {\left\{ {{{\bf{w}}_i}} \right\},{\bm\Phi}} \right)}{\sum_{k=1}^K \|\mathbf{w}_k\|^2 + P_{\rm C}} \label{p1:2a2}\\
    \rm{s.t.}~& 0 \le x_{n,m}^{\rm P} \le D, ~\forall n,m, \label{p1:2b} \\
    & x_{n,m}^{\rm P} - x_{n,m-1}^{\rm P} \ge \Delta_{\min}, ~\forall n,m>1, \label{p1:2c} \\
    & \sum\nolimits_{k=1}^K \|\mathbf{w}_k\|^2 \le P_{\max}, \label{p1:2d} \\
    &\phi_l \in [0,2\pi), ~\forall l, \label{p1:2e} 
\end{align}
\end{subequations}
where ${P_{\rm C}}$ denotes the constant circuit power, $\Delta_{\min} > 0$ denotes the minimum spacing between any two adjacent PAs, and ${P_{\rm max}}$ denotes the total power budget.


\section{Three-Stage GNN for Joint Optimization}

\begin{figure*}[t]
{\centering
{\includegraphics[ width=1\textwidth]{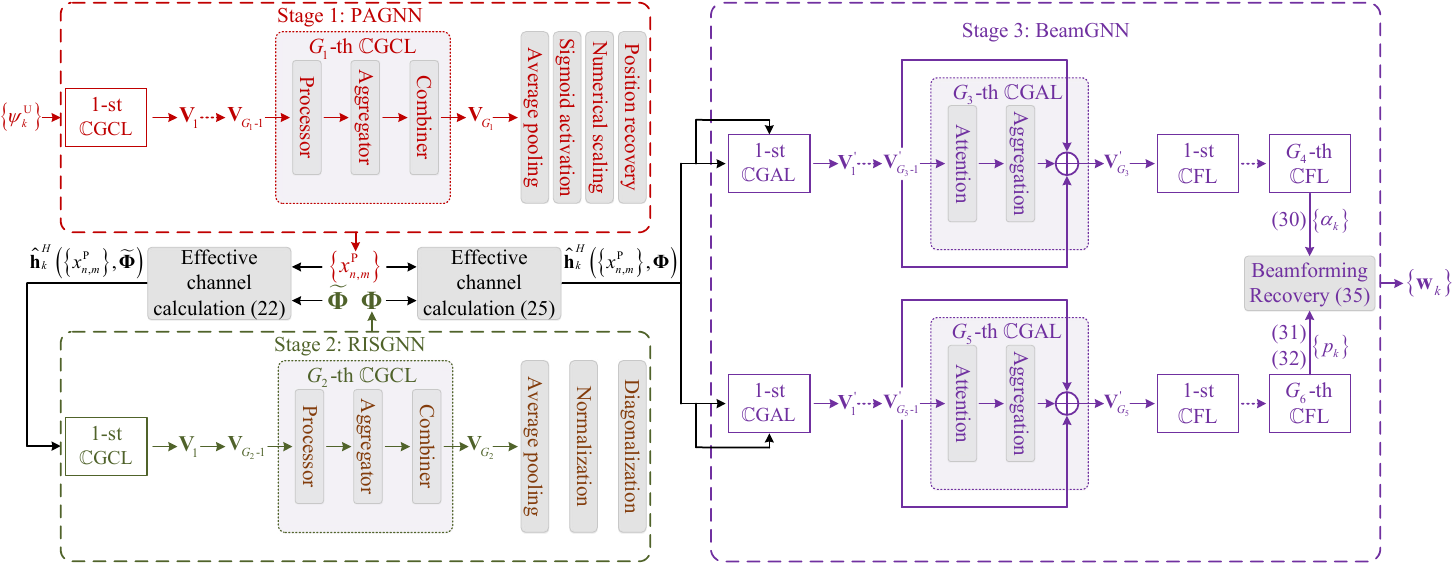}}}
\caption{The structure of the proposed three-stage GNN maps the given user locations $\{\bm{\psi}^{\rm U}_k\}$ to a complete solution $\{x^{\rm P}_{n,m}, \bm{\Phi},\mathbf{w}_k\}$ through three sequential stages. \textbf{Stage 1} (PAGNN) employs $G_1$ ${\mathbb C}$GCL layers to map the user locations to a set of PA positions $\{x^{\rm P}_{n,m}\}$. \textbf{Stage 2} (RISGNN) employs $G_2$ ${\mathbb C}$GCL layers to map the effective channels $\{{\widehat{\bf h}^H_k(\{x^{\rm P}_{n,m}\},\bm{\widetilde{\Phi}})}\}$ computed by \eqref{ec_s2} to a RIS phase shift matrix $\bm{\Phi}$. \textbf{Stage 3} (BeamGNN) uses two parallel branches to learn beamforming parameters: after updating the $\{{\widehat{\bf h}^H_k(\{x^{\rm P}_{n,m}\},{\bm\Phi})}\}$ by \eqref{effective_channel}, the top (bottom) branch employs $G_3$ $\mathbb{C}$GAL ($G_5$ $\mathbb{C}$GAL) layers and $G_4$ $\mathbb{C}$FL ($G_6$ $\mathbb{C}$FL) layers to map the obtained $\{{\widehat{\bf h}^H_k(\{x^{\rm P}_{n,m}\},{\bm\Phi})}\}$ to a set of hybrid coefficients $\{\alpha_k\}$ (power allocation $\{p_k\}$). The obtained $\{p_k\}$ and $\{\alpha_k\}$ are then utilized to yield the output beamforming vectors $\{\mathbf{w}_k\}$.}
\label{structure}
\end{figure*}

To leverage the graph-structured topology of the considered system while handling Problems P1 and P2, we convert it into a fully connected directed graph denoted by $\mathcal{G} = (\mathcal{V}, \mathcal{E})$. Here, $\mathcal{V}$ with $|\mathcal{V}| = K$ denotes the set of nodes, each of which represents one user, and $\mathcal{E}$ denotes the set of edges, each of which represents the inter-node  relationship. Our goal is to construct and train a DL model to map the graph to a near-optimal solution to Problems P1 and P2.

\subsection{Overall Framework}

We design a three-stage GNN as illustrated in Figure \ref{structure}. Our proposed GNN consists of three sequential yet interdependent sub-models, i.e, 1) PAGNN for  optimizing PA positions, 2) RISGNN for tuning RIS phase shifts, 3) BeamGNN for designing baseband beamformers. Particularly, the three sub-models realize the following functions, respectively:
\begin{itemize}
    \item \textbf{PAGNN (Stage 1)} learns the PA positions $\{\bm{\psi}_{n,m}^{\rm P}\}$ from user locations $\{\bm{\psi}^{\rm U}_k\}$;
    
    \item \textbf{RISGNN (Stage 2)} learns the RIS phase shift matrix ${\bm\Phi}$ from  effective channels computed with the obtained PA positions $\{\bm{\psi}_{n,m}^{\rm P}\}$ and an initial phase-shift matrix $\bm{\widetilde{\Phi}}$;
    
    \item \textbf{BeamGNN (Stage 3)} learns baseband beamforming vectors $\{\mathbf{w}_k\}$ from  effective channels computed with both the obtained PA positions $\{\bm{\psi}_{n,m}^{\rm P}\}$ and the phase shifts ${\bm\Phi}$.
\end{itemize}

All sub-models are built upon a GNN backbone, incorporating layers chosen from the  complex graph convolution layer (${\mathbb C}$GCL),  complex graph attention layer (${\mathbb C}$GAL), and  complex fully-connected layer (${\mathbb C}$FL). 

In the following subsections, we detail the structure and operations of each stage, and then present the detailed operations of the three layers.

\subsection{Stage 1: PAGNN}

The input and the output of PAGNN are  $\{\bm{\psi}^{\rm U}_k\}$ and $\{x^{\rm P}_{n,m}\}$, respectively. Node $k$ is initialized with a feature of $\bm{\psi}^{\rm U}_k$.

\subsubsection{Feasibility Guaranteed for PA Placement}
To guarantee that the learned PA positions $\{x^{\rm P}_{n,m}\}$ satisfy constraints \eqref{p1:2b} and \eqref{p1:2c}, we design a feasibility guaranteed module with the following  auxiliary variables:
\begin{flalign}
\delta_{n,1} = x^{\rm P}_{n,1}, \delta_{n,m} = x^{\rm P}_{n,m} - x^{\rm P}_{n,m-1} - \Delta_{\min}, \forall n,m>1,
\end{flalign}
where $\delta_{n,m}$ represents the spacing between adjacent antennas. Then, the maximum available spacing is given by 
\begin{flalign}
\delta_{\max} = D - (M-1)\Delta_{\min}.
\end{flalign}
Constraints \eqref{p1:2b} and \eqref{p1:2c} can be equivalently expressed as 
\begin{flalign}\label{cons}
\delta_{n,m} \geq 0,~\sum\nolimits_{m^{\prime}=1}^{M} \delta_{n,m^{\prime}} \leq \delta_{\max},\forall n,m.
\end{flalign}

 \subsubsection{${\mathbb C}$GCL in PAGNN} Thus, we employ $G_1$ ${\mathbb C}$GCLs to construct the mapping from  $\{\bm{\psi}^{\rm U}_k\}$ to $\{\delta_{n,m}\}$. Denote the output of the $G_1$-th ${\mathbb C}$GCL as $\mathbf{V}_{G_1} \in \mathbb{C}^{K \times (N \cdot M)}$. Since PA positions are graph-level outputs rather than node-specific features, we apply average pooling across all $K$ nodes: 
\begin{flalign}
{\overline \delta}_{n,m} = \frac{1}{K} \sum\nolimits_{k=1}^{K} \text{Re}\left(\left[\mathbf{V}_{G_1}\right]_{k,(n-1)M+m}\right).
\end{flalign}
To ensure constraint \eqref{cons} satisfied, we apply the sigmoid activation $\sigma(\cdot)$ and numerical scaling to obtain
\begin{flalign}
\delta_{n,m} = \delta_{\max} \times \sigma({\overline \delta}_{n,m}),
\end{flalign}
\begin{flalign}
\delta_{n,m} := \left\{ {\begin{array}{*{20}{l}}
{{\delta_{n,m}},~\sum_{i=1}^{M}  {{\delta_{n,i}}}  \le {\delta_{\max }}}\\
{\frac{{{\delta_{n,m}}}}{{\sum_{i=1}^{M}  {{\delta_{n,i}}} }}\delta_{\max },~\sum_{i=1}^{M}  {{\delta_{n,i}}}  > {\delta_{\max }}}
\end{array}} \right..
\end{flalign}

Subsequently, PA positions can be recovered by 
\begin{flalign}
x^{\rm P}_{n,m} = \sum\nolimits_{i=1}^{m} \delta_{n,i}+(m-1)\Delta_{\min}, \forall n,m>1.
\end{flalign}
This procedure ensures that the output can satisfy constraints \eqref{p1:2b} and \eqref{p1:2c}.

\subsection{Stage 2: RISGNN}

The input and output of  RISGNN are effective channels with an initialized RIS phase shift matrix  (cf. \eqref{ec_s2} below) and the RIS phase shift matrix ${\bm\Phi}$, respectively.

\subsubsection{Pre-processing}

With the obtained  $\{\bm{\psi}^{\rm P}_{n,m}\}$, we calculate the pinching beamforming matrix $\mathbf{G}(\{x_{n,m}^{\rm P}\}) $ following \eqref{eqg1} and \eqref{eqg2}. Besides, the PA-user channel $\mathbf{f}_k(\{x_{n,m}^{\rm P}\})$ is computed via \eqref{direct channel1} and \eqref{direct channel2}, and the PA-RIS channel $\mathbf{H}(\{x_{n,m}^{\rm P}\})$ is calculated through \eqref{ris channel1}-\eqref{ris channel4}, while the RIS-user  channel $\mathbf{h}_k$ is pre-defined. Moreover, we initialize the RIS phase shift matrix as $\bm{\widetilde{\Phi}} = \text{diag}\{[1, 1, \ldots, 1]\} \in \mathbb{C}^{L \times L}$. Then, the effective channel in Stage 2 is defined by 
\begin{flalign}\label{ec_s2}
&{\widehat{\bf h}^H_k\left(\left\{x^{\rm P}_{n,m}\right\}, \bm{\widetilde{\Phi}} \right)}=\\
&\left( {{\bf f}_k^H\left(\left\{{x_{n,m}^{\rm P}}\right\}\right)+{\bf{h}}_{k}^H\bm{\widetilde{\Phi}}  {\bf H}\left(\left\{{x_{n,m}^{\rm P}}\right\}\right)} \right){\bf G}\left(\left\{{x_{n,m}^{\rm P}}\right\}\right),\nonumber
\end{flalign}
which is input into RISGNN as  the feature of node $k$.

\subsubsection{${\mathbb C}$GCL in RISGNN}
We employ $G_2$ ${\mathbb C}$GCLs to construct the mapping from  $\{{\widehat{\bf h}^H_k(\{x^{\rm P}_{n,m}\},\bm{\widetilde{\Phi}})}\}$ to  $\bm{{\Phi}}$. Similar to Stage 1, $\bm{{\Phi}}$ is a graph-level output shared by all nodes.

Denote the output of the $G_2$-th ${\mathbb C}$GCL as $\mathbf{V}_{G_2} \in \mathbb{C}^{K \times L}$, and we apply average pooling across all the $K$ nodes to obtain
\begin{flalign}
\psi_l = \frac{1}{K} \sum\nolimits_{k=1}^{K} [\mathbf{V}_{G_2}]_{k,l} \in \mathbb{C}.
\end{flalign}
 To satisfy constraint \eqref{p1:2e}, $\psi_l $ is normalized  by
\begin{flalign}
e^{j\phi_l}=\psi_l := \frac{\psi_l}{|\psi_l|}, \forall l,
\end{flalign}
ensuring $|e^{j\phi_l}| = 1$.
The  RIS phase shift matrix $\bm\Phi$ is constructed following \eqref{rismatrix}.

\subsection{Stage 3: BeamGNN}

The input and output of  RISGNN are effective channels with the obtained RIS phase shift matrix in Stage 2 and baseband beamforming vectors $\{{\bf w}_k\}$, respectively.


\subsubsection{Pre-processing}
With the obtained ${\bm\Phi}$, we update the effective channel in Stage 3 as
\begin{flalign}\label{effective_channel}
&{\widehat{\bf h}^H_k\left(\left\{x^{\rm P}_{n,m}\right\},{\bm\Phi}\right)} :=\\
&\left( {{\bf f}_k^H\left(\left\{{x_{n,m}^{\rm P}}\right\}\right)+{\bf{h}}_{k}^H\bm{{\Phi}}  {\bf H}\left(\left\{{x_{n,m}^{\rm P}}\right\}\right)} \right){\bf G}\left(\left\{{x_{n,m}^{\rm P}}\right\}\right),\nonumber
\end{flalign}
which is input into BeamGNN as  the feature of node $k$.

\subsubsection{HZM Learning}
To mitigate complexity of direct learning beamforming vectors, we adopt a model-based method, termed hybrid zero-forcing and maximum ratio transmission (HZM) learning \cite{hyb}. This approach decomposes each beamforming vector into its power part and its direction part as
\begin{flalign}\label{HZM}
\mathbf{w}_k = \sqrt{p_k} \overline{\mathbf{w}}_k(\alpha_k), ~ \|\overline{\mathbf{w}}_k(\alpha_k)\|^2 = 1,
\end{flalign}
where $p_k \in \mathbb{R}^+$ denotes the transmit power allocated to user $k$, and $\overline{\mathbf{w}}_k(\alpha_k) \in \mathbb{C}^N$ is the unit-norm direction vector parameterized by a hybrid coefficient $\alpha_k \in [0,1]$, which  is computed as
\begin{flalign}
\overline{\mathbf{w}}_k(\alpha_k) = \frac{\alpha_k \frac{\mathbf{u}_k}{\|\mathbf{u}_k\|} + (1-\alpha_k) \frac{{\widehat{\bf h}_k\left(\left\{x^{\rm P}_{n,m}\right\},{\bm\Phi}\right)}}{\|{\widehat{\bf h}_k\left(\left\{x^{\rm P}_{n,m}\right\},{\bm\Phi}\right)}\|}}{\left\|\alpha_k \frac{\mathbf{u}_k}{\|\mathbf{u}_k\|} + (1-\alpha_k) \frac{{\widehat{\bf h}_k\left(\left\{x^{\rm P}_{n,m}\right\},{\bm\Phi}\right)}}{\|{\widehat{\bf h}_k\left(\left\{x^{\rm P}_{n,m}\right\},{\bm\Phi}\right)}\|}\right\|},
\end{flalign}
where $\mathbf{u}_k$ is the $k$-th column of the zero-forcing matrix $\mathbf{U} \in \mathbb{C}^{N \times K}$ given by
\begin{flalign}
\mathbf{U} = \mathbf{Z}^H(\mathbf{Z}\mathbf{Z}^H)^{-1},
\end{flalign}
where 
\begin{flalign}
\mathbf{Z} = &\left[{\widehat{\bf h}^H_1\left(\left\{x^{\rm P}_{n,m}\right\},{\bm\Phi}\right)}; {\widehat{\bf h}^H_2\left(\left\{x^{\rm P}_{n,m}\right\},{\bm\Phi}\right)}; \right.\\&\left.\ldots, \widehat{\bf h}^H_K\left(\left\{x^{\rm P}_{n,m}\right\};{\bm\Phi}\right)\right]\in \mathbb{C}^{K \times N}. \nonumber 
\end{flalign}
Notably, the HZM learning reduces the required output dimension from $(K\times N)$ complex values to $2K$ real values, which can enhance both the expressiveness and training efficiency of the BeamGNN.

 \subsubsection{${\mathbb C}$GAL and $\mathbb{C}$FL in BeamGNN}

We employ two parallel branches to separately learn hybrid coefficients and power allocation.

For the hybrid coefficients, we use $G_3$ $\mathbb{C}$GAL layers followed by $G_4$ $\mathbb{C}$FL layers to construct the mapping from $\{{\widehat{\bf h}^H_k\left(\left\{x^{\rm P}_{n,m}\right\},{\bm\Phi}\right)}\}$ to $\{\alpha_k\}$. Denote the output of the $G_4$-th $\mathbb{C}$FL layer as $\mathbf{V}_{G_4} \in \mathbb{C}^{K}$. We apply the sigmoid activation function to ensure $\alpha_k \in (0,1)$:
\begin{flalign}
\alpha_k = \sigma(\text{Re}([\mathbf{V}_{G_4}]_{k})),~\forall k.
\end{flalign}

For the power allocation, we use $G_5$ $\mathbb{C}$GAL layers followed by $G_6$ $\mathbb{C}$FL layers to construct the mapping from $\{{\widehat{\bf h}^H_k\left(\left\{x^{\rm P}_{n,m}\right\},{\bm\Phi}\right)}\}$ to $\{p_k\}$. Denote the output of the $G_6$-th $\mathbb{C}$FL layer as $\mathbf{V}_{G_6} \in \mathbb{C}^{K}$. We obtain preliminary power value within $(0, P_{\max})$ as
\begin{flalign}
\widetilde{p}_k = P_{\max} \times \sigma\left(\text{Re}\left(\left[\mathbf{V}_{G_6}\right]_{k}\right)\right),~\forall k.
\end{flalign}

\subsubsection{Feasibility Guaranteed for Beamforming Vectors}

To guarantee that the learned power allocation satisfies  constraint \eqref{p1:2d}, we apply a customized normalization activation function:
\begin{flalign}
p_k := \left\{ {\begin{array}{*{20}{l}}
{\widetilde{p}_k,~\sum\nolimits_{k=1}^K  {\widetilde{p}_i}  \le {P_{\max }}}\\
{\frac{{\widetilde{p}_k}}{{\sum\nolimits_{k=1}^K  {\widetilde{p}_i} }}P_{\max },~\sum\nolimits_{k=1}^K  {\widetilde{p}_i}  > {P_{\max }}}
\end{array}} \right..
\end{flalign}
Consequently, we have $\sum\nolimits_{k=1}^K \|\mathbf{w}_k\|^2 = \sum\nolimits_{k=1}^K p_k \leq P_{\max}$, ensuring constraint \eqref{p1:2d} satisfied.

Finally, the desired beamforming vectors can be recovered following \eqref{HZM} with the obtained $\{\alpha_k\}$ and  $\{p_k\}$.

\subsection{Detailed Processes of ${\mathbb C}$GCL,  ${\mathbb C}$GAL, and ${\mathbb C}$FL}

In this subsection, we detail the three types of layers (i.e.,  ${\mathbb C}$GCL, the $\mathbb{C}$GAL, and  $\mathbb{C}$FL) employed in our proposed three-stage model. These layers are tailored to handle complex-valued wireless channel features while capturing the graph-structured relationships between users.

\subsubsection{Complex Graph Convolution Layer}

The ${\mathbb C}$GCL is specifically designed to extract position-related and channel-related features through message passing on the constructed graph.  This layer is particularly effective for learning PA positions (cf. PAGNN) and RIS phase shift coefficients (cf. RISGNN).

The ${\mathbb C}$GCL enables each node to aggregate information from its neighbors via a processor-aggregator-combiner architecture. Consider the $g$-th ${\mathbb C}$GCL with input node features $\mathbf{V}_{g-1} \in \mathbb{C}^{K \times D_{g-1}}$, where $D_{g-1}$ denotes the feature dimension. For a given node $k$, the $g$-th ${\mathbb C}$GCL computes pairwise interference features between itself and its neighbors through a processor function $q(\cdot, \cdot)$:
\begin{flalign}
\left[\mathbf{M}_{g}\right]_{k,k^{\prime},:} = q\left(\left[\mathbf{V}_{g-1}\right]_{k,:}, \left[\mathbf{V}_{g-1}\right]_{k^{\prime},:}\right),
\end{flalign}
where $\mathbf{M}_{g} \in \mathbb{C}^{K \times K \times J_{g}}$ represents the processed message from node $k^{\prime}$ to node $k$, and $J_{g}$ denotes the hidden dimension.

The messages from all neighbors are then aggregated through summation:
\begin{flalign}
\left[\mathbf{T}_{g}\right]_{k,:} = \sum\nolimits_{k^{\prime}=1}^K \left[\mathbf{M}_{g}\right]_{k,k^{\prime},:},
\end{flalign}
where self-loops are included to allow each node to retain its own information. Subsequently, $[\mathbf{T}_{g}]_{k,:}$ is combined with the node's own feature $[\mathbf{V}_{g-1}]_{k,:}$ through a combiner function $f(\cdot, \cdot)$:
\begin{flalign}
\left[\mathbf{V}_g\right]_{k,:} = f\left(\left[\mathbf{V}_{g-1}\right]_{k,:}, \left[\mathbf{T}_{g}\right]_{k,:}\right).
\end{flalign}
The processor $q(\cdot, \cdot)$ and the combiner $f(\cdot, \cdot)$ are implemented via separate  MLPs.

\subsubsection{Complex Graph Attention Layer}

The $\mathbb{C}$GAL is tailored  to learn node-level beamforming patterns by capturing inter-user interactions via multi-head attention (cf. BeamGNN). Specifically, for a given node, its attention coefficients are adaptively assigned to its neighbors according to their importance.


\begin{figure*}
\begin{flalign}
\label{attention coefficient}
&[{\bf{A}}_{g}]_{k,k^{\prime}} = 
\frac{
    \exp \left( 
        {{\rm{{\mathbb R}LeakyReLU}}\left( 
            {{\mathop{\rm Re}\nolimits} \left( 
                {\bf{a}}_{g}^T
                \left( 
                    {\rm{Concat}}\left( 
                        {[\mathbf{V}^{\prime}_{(g - 1)}]_{k,:} \mathbf{W}_{g}}, 
                        {[\mathbf{V}^{\prime}_{(g - 1)}]_{k^{\prime},:} \mathbf{W}_{g}} 
                    \right) 
                \right) 
            \right)} 
        \right)} 
    \right)
}{
    \sum\nolimits_{k^{\prime\prime}=1}^K 
    \exp \left( 
        {{\rm{{\mathbb R}LeakyReLU}}\left( 
            {{\mathop{\rm Re}\nolimits} \left( 
                {\bf{a}}_{g}^T
                \left( 
                    {\rm{Concat}}\left( 
                        {[\mathbf{V}^{\prime}_{(g - 1)}]_{k,:} \mathbf{W}_{g}}, 
                        {[\mathbf{V}^{\prime}_{(g - 1)}]_{k^{\prime\prime},:} \mathbf{W}_{g}} 
                    \right) 
                \right) 
            \right)} 
        \right)} 
    \right)
}
\end{flalign}
\hrule
\end{figure*}

Consider the $g$-th $\mathbb{C}$GAL with input node features $\mathbf{V}^{\prime}_{g-1} \in \mathbb{C}^{K \times D^{\prime}_{g-1}}$, where $D^{\prime}_{g-1}$ denotes the corresponding feature dimension. The attention coefficient from node $k^{\prime}$ to node $k$ is given by \eqref{attention coefficient}, where $\mathbf{W}_g \in \mathbb{C}^{D^{\prime}_{g-1} \times D_g^{\prime}}$ presents the learnable transformation matrix, $\mathbf{a}_g \in \mathbb{C}^{2D_g^{\prime}}$ presents the learnable attention vector, and ${\rm {\mathbb R}LeakyReLU}(\cdot)$ represents the real LeakyReLU activation function.

The updated node features are generated by aggregating the transformed neighbor features weighted by the attention coefficients:
\begin{flalign}
\left[\mathbf{V}^{\prime}_g\right]_{k,:} = \mathbb{C}\text{ReLU}\big(\sum\limits_{k^{\prime}=1}^K \left[{\bf{A}}_{g}\right]_{k,k^{\prime}} \left[\mathbf{V}^{\prime}_{g-1}\right]_{k^{\prime},:} \mathbf{W}_g\big),
\end{flalign}
where $\mathbb{C}\text{ReLU}(\cdot)$ represents the complex ReLU activation function.

To enhance training stability and mitigate over-smoothing in deep GNNs, residual connections are incorporated between consecutive layers. Specifically,
\begin{flalign}
\left[\mathbf{V}^{\prime}_g\right]_{k,:} := \left[\mathbf{V}^{\prime}_g\right]_{k,:} + \left[\mathbf{V}^{\prime}_{g-1}\right]_{k,:} \overline{\mathbf{W}}_g,
\end{flalign}
where $\overline{\mathbf{W}}_g \in \mathbb{C}^{D^{\prime}_{g-1} \times D^{\prime}_g}$ is the learnable residual matrix. 

\subsubsection{Complex Fully-Connected Layer}

The $\mathbb{C}$FL is designed to project the graph-structured features extracted by the $\mathbb{C}$GALs onto the space of the desired beamforming vectors (cf. BeamGNN). The $\mathbb{C}$FL employs standard feedforward neural network operations specifically adapted for complex-valued features.

Consider the $f$-th layer with input features $\mathbf{V}^{\prime\prime}_{f-1} \in \mathbb{C}^{K \times D^{\prime\prime}_{f-1}}$, where $D^{\prime\prime}_{f-1}$ denotes the corresponding feature dimension. Then,  $\mathbf{V}^{\prime\prime}_{f-1}$ is updated by
\begin{flalign}
\mathbf{V^{\prime\prime}}_f = \mathbb{C}\text{ReLU}\left(\mathbf{V}^{\prime\prime}_{f-1} \mathbf{W}^{\prime}_f + \mathbf{B}_f\right),
\end{flalign}
where $\mathbf{W}^{\prime}_f \in \mathbb{C}^{D^{\prime\prime}_{f-1} \times D^{\prime\prime}_f}$ and $\mathbf{B}_f \in \mathbb{C}^{K \times D^{\prime\prime}_f}$ are the learnable weight matrices. 

To ensure scalability to varying numbers of users $K$, all $K$ row vectors of $\mathbf{B}_f$ are set identical, such that $\mathbb{C}$FLs can generalize to unseen problem sizes (i.e., new user numbers) during inference. Each $\mathbb{C}$FL is followed by a complex batch normalization (BN) layer \cite{complex NN} to mitigate  overfitting and improve convergence dynamics.

\subsection{Unsupervised Loss Function}

For a given input $\{\bm{\psi}^{\rm U}_k\}$, the proposed model generates  the complete solution, i.e, $\{\mathbf{w}_k, \{x^{\rm P}_{n,m}\}, {\bm\Phi}\}$ in a joint optimization approach, by sequentially executing its three stages. The three-stage GNN model can be trained in an end-to-end unsupervised manner by directly optimizing the system utility, since all constraints are guaranteed to be satisfied. 


Denote all learnable parameters of the three-stage GNN as $\bm \Theta$, which encompasses parameters of all ${\mathbb C}$GCLs, $\mathbb{C}$GALs, and $\mathbb{C}$FLs across the three stages. Crucially, the dimensionality of $\bm \Theta$ is independent of the number of users $K$, thus ensuring the trained model can be directly applied to problem instances with varying user numbers without retraining. Moreover, the three-stage GNN can be trained to solve both Problems P1 and P2 in a unified framework, with the only difference being the adoption of different loss functions.

We directly minimize the reciprocal of each objective function. For Problem P1, the loss function over a mini-batch of size $T$ is defined as
\begin{flalign}\label{loss1}
\mathcal{L}_T(\bm \Theta) = \frac{1}{T} \sum_{t=1}^{T} \frac{1}{\sum_{k=1}^{K} R_k^{(t)}(\{\{\mathbf{w}_k\}, \{x^{\rm P}_{n,m}\}, {\bm\Phi}|\bm \Theta\})}.
\end{flalign}
For Problem P2, the loss function is defined as
\begin{flalign}\label{loss2}
\mathcal{L}_T(\bm \Theta) = \frac{1}{T} \sum_{t=1}^{T} \frac{\sum_{k=1}^{K} \|\mathbf{w}_k^{(t)}\|^2 + P_{\rm C}}{\sum_{k=1}^{K} R_k^{(t)}(\{\{\mathbf{w}_k\}, \{x^{\rm P}_{n,m}\}, {\bm\Phi}|\bm \Theta\})}.
\end{flalign}

\section{Implementation Strategies of Three-Stage GNN}

\begin{figure*}[t]
\begin{flalign}
\mathbf{w}_k = \sqrt{p_k} \frac{\left(\mathbf{I} + \sum_{k'=1}^K \frac{\lambda_{k'}}{\sigma^2_k} {\widehat{\bf h}_{k^{\prime}}\left(\left\{x^{\rm P}_{n,m}\right\},{\bm\Phi}\right)} {\widehat{\bf h}^H_{k^{\prime}}\left(\left\{x^{\rm P}_{n,m}\right\},{\bm\Phi}\right)} \right)^{-1} {\widehat{\bf h}_{k}\left(\left\{x^{\rm P}_{n,m}\right\},{\bm\Phi}\right)}}{\left\| \left(\mathbf{I} + \sum_{k'=1}^K \frac{\lambda_{k'}}{\sigma^2_k} {\widehat{\bf h}_{k^{\prime}}\left(\left\{x^{\rm P}_{n,m}\right\},{\bm\Phi}\right)} {\widehat{\bf h}^H_{k^{\prime}}\left(\left\{x^{\rm P}_{n,m}\right\},{\bm\Phi}\right)} \right)^{-1} \widehat{\bf h}_{k}\left(\left\{x^{\rm P}_{n,m}\right\},{\bm\Phi}\right) \right\|}
\label{eq:rzf}
\end{flalign}
\hrule
\end{figure*}

\begin{figure}[t]
{\centering
{\includegraphics[ width=.48\textwidth]{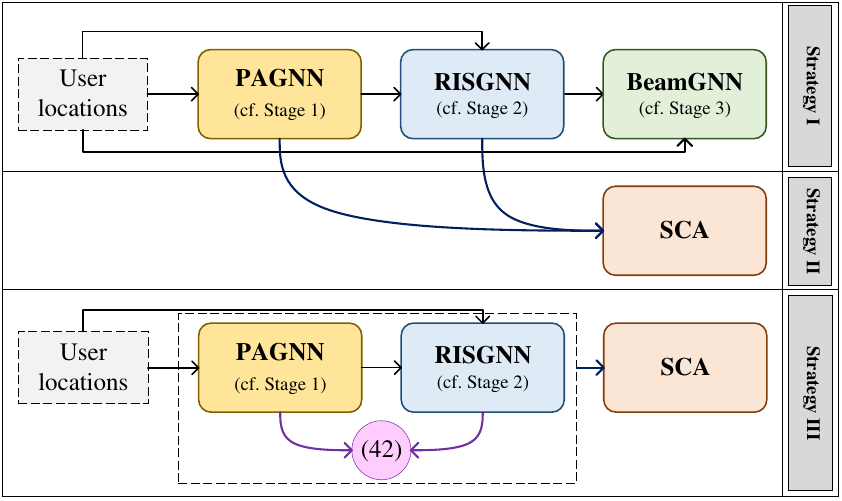}}}
\caption{Illustration of three strategies.} 
\label{3strategies}
\end{figure}

Conventional optimization methods can efficiently optimize beamforming vectors to obtain stationary-point solutions when  PA positions and the RIS phase shift matrix are pre-defined; yet, jointly optimizing these parameters remains challenging. The outputs of PAGNN and RISGNN can be used as prior information for optimization methods. Next, let us consider three implementation strategies, as illustrated in Figure \ref{3strategies} for the three-stage GNN in cases with/without convex optimization methods involved, to understand their efficacy in terms of computational efficiency, optimality, and complexity, so that the proposed GNN can offer a practical solution to the transmission design for the RIS-assisted PASS.



\subsection{Strategy I: Fully Learning-Based Approach}

Strategy I, detailed in Section IV, employs the full three-stage GNN architecture, with all decision variables, i.e., PA positions $\{x_{n,m}^{\rm P}\}$, RIS phase shift matrix $\bm\Phi$, and beamforming vectors $\{\mathbf{w}_k\}$, directly output as a complete solution set.

This end-to-end learning paradigm delivers the fastest inference speed, as all three stages perform forward passes through the neural network, eliminating the need for iterative optimization procedures. However, the learned beamforming vectors may suffer from a performance loss compared to stationary-point solutions, particularly when the test scenarios deviate substantially  from the training scenarios.



\subsection{Strategy II: Hybrid Learning-Optimization Approach}

Strategy II leverages the full three-stage GNN architecture to learn only the PA positions $\{x_{n,m}^{\rm P}\}$ and RIS phase shift matrix $\bm\Phi$ yielded in the first two stages, respectively. Then, with the obtained $\{x_{n,m}^{\rm P}\}$ and $\bm\Phi$, the effective channels are computed using \eqref{effective_channel} and fed into an SCA-based algorithm to obtain the desired beamforming vectors. Notably, the second strategy leverages the full three-stage GNN architecture to learn only the PA positions $\{x_{n,m}^{\rm P}\}$ and RIS phase shift matrix $\bm\Phi$. The beamforming vectors, while indeed learned, are not utilized.  Strategy II can provide a trade-off between inference speed and solution optimality, while Strategy I primarily focuses on the inference time for a practical solution.






\subsection{Strategy III: Model-Based Hybrid Learning-Optimization Approach}


Strategy III employs a closed-form expression \eqref{eq:rzf} (where $p_k = \lambda_k = P_{\max}/K$) \cite{bjornson2014optimal} in \eqref{loss1} or \eqref{loss2} to alleviate the usage of BeamGNN, and then employs the SCA-based algorithm (as by Strategy II) to obtain the desired $\{{\bf w}_k\}$. In this case, the three-stage GNN is simplified as two-stage GNN (comprising PAGNN and RISGNN) to directly yield the PA positions $\{x_{n,m}^{\rm P}\}$ and RIS phase shift matrix $\bm\Phi$, thereby saving  inference time than Strategy II. 





\section{NUMERICAL RESULTS}

This section provides the numerical results to evaluate our proposed three-stage GNN as well as its three implementation strategies. The architecture of three-stage GNN under test is given in Table \ref{PAGNNstructure}.

\begin{table}[t]
    \centering
    \footnotesize
    \caption{The architecture information of the proposed three-stage GNN under test.}
    \label{PAGNNstructure}
    \begin{tabular}{c||c|c|c|c|c|c}
      \hline
      {\#Stage} & {\#Layer} & No. & {Type} & {IFs} & {OFs} & {BN}\\
      \hline
      \hline
      \multirow{3}{*}{\makecell{1\\(PAGNN)}} & \multirow{3}{*}{$G_1=3$} &1 & ${\mathbb C}$GCL & 3 & 1,024 & $\times$\\
      \cline{3-7}
      & &2 & ${\mathbb C}$GCL & 1,024 & 1,024 & $\times$ \\
      \cline{3-7}
      & &3 & ${\mathbb C}$GCL & 1,024 & $M \times N$ & $\times$ \\
      \hline
      \multirow{3}{*}{{\makecell{2\\(RISGNN)}}} & \multirow{3}{*}{$G_2=3$} &4 & ${\mathbb C}$GCL & $N$ & 1,024 & $\times$\\
      \cline{3-7}
      & &5 & ${\mathbb C}$GCL & 1,024 & 1,024 & $\times$ \\
      \cline{3-7}
      & &6 & ${\mathbb C}$GCL & 1,024 & $L$ & $\times$ \\
      \hline
      \multirow{8}{*}{{\makecell{3a\\(BeamGNN\\for $\{\alpha_k\}$)}}} & \multirow{3}{*}{$G_3=3$} &7 & ${\mathbb C}$GAL & $N$ & 1,024 & $\times$\\
      \cline{3-7}
      & &8 & ${\mathbb C}$GAL & 1,024 & 1,024 & $\times$ \\
      \cline{3-7}
      & &9 & ${\mathbb C}$GAL & 1,024 & 1,024 & $\times$ \\
      \cline{2-7}
      &\multirow{5}{*}{$G_4=5$} &10 & ${\mathbb C}$FL & 1,024 & 1,024 & \checkmark\\
      \cline{3-7}
      & &11 & ${\mathbb C}$FL & 1,024 & 1,024 & \checkmark\\
      \cline{3-7}
      & &12 & ${\mathbb C}$FL & 1,024 & 1,024 & \checkmark\\
      \cline{3-7}
      & &13 & ${\mathbb C}$FL & 1,024 & 512 & \checkmark\\
      \cline{3-7}
      & &14 & ${\mathbb C}$FL & 512 & 1 & $\times$\\
      \hline
      \multirow{8}{*}{{\makecell{3b\\(BeamGNN\\for $\{p_k\}$)}}} & \multirow{3}{*}{{$G_5=3$}} & 13 & ${\mathbb C}$GAL & $N$ & 1,024 & $\times$\\
      \cline{3-7}
      & &15 & ${\mathbb C}$GAL & 1,024 & 1,024 & $\times$ \\
      \cline{3-7}
      & &16 & ${\mathbb C}$GAL & 1,024 & 1,024 & $\times$ \\
      \cline{2-7}
      & \multirow{5}{*}{{$G_6=5$}}& 17 & ${\mathbb C}$FL & 1,024 & 1,024 & \checkmark\\
      \cline{3-7}
      & &18 & ${\mathbb C}$FL & 1,024 & 1,024 & \checkmark\\
      \cline{3-7}
      & &19 & ${\mathbb C}$FL & 1,024 & 1,024 & \checkmark\\
      \cline{3-7}
      & &20 & ${\mathbb C}$FL & 1,024 & 512 & \checkmark\\
      \cline{3-7}
      & &21 & ${\mathbb C}$FL & 512 & 1 & $\times$\\
      \hline
    \end{tabular}
  \begin{tablenotes}
	\footnotesize
	\item IFs: The dimension of the input features.
	\item OFs: The dimension of the output features.
  \end{tablenotes}
\end{table}


\subsection{Simulation Setting}
\subsubsection{Simulation Scenario} The simulation scenario is according to Figure 1. We consider a serving area with length $S\in\{10,20\}$ m and width $D\in\{10,15,20,25,30\}$ m. The numbers of waveguides, users, PAs per waveguide, and RIS reflecting elements are respectively set as $N\in\{2,4,6,8\}$, $K\in\{1,2,3,4,5,6,7\}$,  $M\in\{2,5,8\}$, and $L\in\{16,32\}$. The waveguide height is set as $H=5$ m. The minimum spacing between adjacent PAs is set as $\Delta_{\min}=0.1$ m. The total power budget is set as $P_{\max}=10$ W. The circuit power is set as $P_{\rm C}=5$ W. The noise power at each user is set as $\sigma^2_k=-60$ dBm. The comprehensive simulation parameters are summarized in Table \ref{Simulation Parameters}.



 For the spatial deployment, users are uniformly distributed within the serving region ranging from $0$ to $D$ m along the $x$-axis and from $-S/2$ to $S/2$ m along the $y$-axis. The $N$ waveguides are mounted at a height of $H$, with their $y$-coordinates spaced evenly across the range 
$[-S/2,S/2]$. The $y$-coordinate of the $n$-th waveguide is given by
\begin{flalign}
    y^{\rm P}_n = (n-1)\times \frac{S}{N} - (N-1) \times \frac{S}{2N}.
\end{flalign}
The RIS is deployed at a fixed location $\psi^{\rm R} = (D/2, 0, H/2)$.

\begin{table}[t]
\footnotesize
\centering
\caption{Simulation Parameters Used in Simulation.}
\label{Simulation Parameters}
\begin{tabular}{c|c}
\hline
{\bf Notation} & {\bf Values} \\ \hline
\hline
Number of waveguides & $N\in\{2,4,6,8\}$ \\ \hline
Number of users & $K\in\{1,2,3,4,5,6,7\}$\\ \hline
Number of PAs per waveguide & $M \in \{2,5,8\}$\\ \hline
Number of RIS reflecting elements & $L \in \{16,32\}$ \\ \hline
Power budget & $P_{\max} = 10$ W \\ \hline
Circuit power & $P_{\rm C} = 5$ W \\ \hline
Noise power & $\sigma^2_k = -60$ dBm \\\hline
Serving area size & $S \times D \in \{10\times10, 20\times20\}$ m$^2$\\ \hline
Waveguide height & $H = 5$ m \\ \hline
Free-space carrier frequency & $f_{\rm c}=6$ GHz \\ \hline
Free-space carrier wavelength & $\lambda = 3\times10^8/f_{\rm c} = 0.05$ m\\ \hline
Minimum inter-PA spacing & $\Delta = \lambda /2$\\ \hline
Minimum PA spacing & $\Delta_{\min} = 0.1$ m \\ \hline
Effective refractive index & $n_{\rm eff}=1.4$ \\\hline
Rician factor & $\kappa=3$ dB \\\hline
Fading exponent & $\alpha=2.8$ \\\hline
Channel gain at 1 m & $\beta_0=-20$ dB \\\hline
\end{tabular}
\end{table}

\subsubsection{Baselines} 

The considered RIS-assisted PASS is abbreviated as {\emph{RIS+PA}}. To numerically evaluate the proposed three-stage GNN as well as the effectiveness of RIS and PAs, the following models and system configurations are considered.

\textbf{Model baselines:}
\begin{itemize}
\item {\emph{CVX}}: An SCA-based algorithm to optimize beamforming vectors with pre-given PA positions and RIS reflecting shift matrix, which is implemented via the MATLAB CVX Toolbox with the MOSEK solver. The convergence tolerance is set to $10^{-4}$. 
\item {\emph{MLP}}: A basic feed-forward MLP with HZM learning to directly map user locations to desired solutions (including $\{\mathbf{w}_k\}, \{x^{\rm P}_{n,m}\}$ and/or ${\bm\Phi}$).
\end{itemize}

\textbf{System baselines:}
\begin{itemize}
\item {\emph{PA-only}}: The case without RIS assistance can be handled via any of four methods: a two-stage model that integrates PAGNN and BeamGNN (following Strategy I); a two-stage model adopting SCA for beamforming optimization (following Strategy II); a combined approach of PAGNN and SCA (following Strategy III); MLP.
\item {\emph{Fixed PA-only}}: The case with PA positions fixed and without RIS assistance, where  PAs of one waveguide are fixed uniformly along it at a spacing of $\Delta_{\min}$,  with their central point at $(D/2,0,H)$. This case can be addressed via three methods: BeamGNN (following Strategy I); CVX (following Strategy II and Strategy III); MLP. Notably, this case can also be regarded as conventional fixed-position antenna systems.
\end{itemize}

\subsubsection{Computer Configuration} 
The DL models are trained and tested under Python 3.10.18 with PyTorch 1.11.0 on a computer with Intel(R) Xeon(R) Gold 6278C CPU and NVIDIA Tesla T4 GPU (16 GB of memory).

\subsubsection{Training and dataset} 
All learnable parameters are initialized using the Kaiming normal initialization method \cite{kaiming_normal}, with the initial learning rate set to $10^{-3}$. A multi-step learning rate scheduler is employed to adaptively decrease the learning rate during training. The Adam optimizer \cite{Adam} is utilized for gradient-based parameter updates during training. Training runs for $100$ epochs with a batch size of $1,024$ samples. To mitigate overfitting, an early stopping mechanism tracks  validation performance, and the parameter set yielding the best validation metric is retained for subsequent testing.

In our experiment, we construct two data categories with distinct purposes. The first one comprises $100,000$ samples in total, split into training, validation, and test subsets in a proportion of 8:1:1. The second one contains $10,000$ samples exclusively for testing to assess generalization performance across unseen configurations. For those performance comparisons involving CVX (cf. Table \ref{Strategy}), we use only $1,000$ test samples per configuration due to the extraordinary computational cost of the CVX solver.



\subsection{Comparison of Three Strategies}

\begin{table}[t]
\footnotesize
\belowrulesep=-0.3pt
\aboverulesep=-0.3pt
\centering
\caption{EE (bit/J/Hz) and SR (bit/s/Hz) comparison of different optimization methods.}
\label{Strategy}
\renewcommand{\arraystretch}{1.15}
\setlength{\tabcolsep}{5pt}
\begin{tabular}{c||cc|cc|cc}
\toprule
\multirow{2}{*}{Method} & 
\multicolumn{2}{c|}{\textbf{RIS+PA}} &
\multicolumn{2}{c|}{\textbf{PA-only}} &
\multicolumn{2}{c}{\textbf{Fixed PA-only}} \\
\cmidrule(lr){2-3} \cmidrule(lr){4-5} \cmidrule(lr){6-7}
& EE & SR & EE & SR & EE & SR \\
\midrule
\textbf{Strategy I} & 8.57 & 63.96 & 7.52 & 57.90 & 6.49 & 51.85 \\
\textbf{Strategy II} & \cellcolor{blue!10}8.62 & \cellcolor{blue!10}65.26 & \cellcolor{blue!10}7.57 & 58.99 & \cellcolor{blue!10}6.53 & \cellcolor{blue!10}52.89 \\
\textbf{Strategy III} & 8.37 & 64.36 & 7.47 & \cellcolor{blue!10}59.22 & \cellcolor{blue!10}6.53 & \cellcolor{blue!10}52.89 \\
\textbf{MLP} & 4.56 & 63.11 & 3.96 & 57.28 & 3.53 & 51.34 \\
\bottomrule
\end{tabular}
\begin{tablenotes}
\footnotesize
\item The best performance for each column is marked in blue. 
\end{tablenotes}
\end{table}

Table \ref{Strategy} compares the SR and EE performance of the previously mentioned three strategies across the configurations of RIS+PA, PA-only, and Fixed PA-only. Besides, we give the results of CVX and MLP, accordingly. 

Among the three strategies, Strategy II delivers the best performance in most configurations. Its superior performance is attributed to its hybrid learning-optimization paradigm: the three-stage GNN rapidly learns near-optimal PA positions and RIS phase shift matrix, while the SCA-based algorithm refines beamforming vectors. Strategy III delivers better performance than Strategy I in four out of six configurations, thanks to the SCA-based algorithm. However, its fixed beamforming structure during learning may induce slight performance degradation in optimizing PA positions and the RIS phase shift matrix compared to Strategy II. Although Strategy I performs worse than Strategies II and III, the performance loss is marginal. Moreover, Strategy I offers the fastest inference speed (as detailed in Table \ref{inference_time}, Subsection VI-E below), as it does not involve the SCA-based algorithm, making it suitable for latency-critical applications.



In a comparison between MLP and Strategy I (both fully learning-based), the proposed GNN-based model exhibits substantial improvements in EE, achieving $87.9\%$, $89.9\%$, and $83.9\%$ better performance than MLP across the three considered configurations, respectively. For SR, the proposed GNN-based model also outperforms MLP, with modest gains across all three configurations.

In the fixed PA-only configuration, Strategy I exhibits a performance gap of less than $0.6\%$ in EE and $2.0\%$ in SR compared to CVX (Strategy II), thus confirming that the learned beamforming vectors are very close to optimal ones, thereby making them more flexible for addressing complex optimization problems.


\subsection{Generalization Across Different Numbers of Users}

Table \ref{tab:generalization} demonstrates the generalization capability of the three-stage GNN across different user numbers. Notably, MLP lacks the generalization capability to unseen user numbers. Such a capability is critical  for practical implementation, as it alleviates the burden of retraining DL models to adapt to dynamic wireless environments. The three-stage GNN, trained on data with  $K_{\text{Tr}}$ users, effectively handles test scenarios with varying $K_{\text{Te}}$ users, exhibiting less than $0.3\%$ performance sensitivity to overlapping configurations. This capability is attributed to the permutation-equivariant property of GNN, which ensures that the learned policies are invariant to node ordering and enables GNN to seamlessly adapt to graphs with varying node numbers. Specifically, the trends for achievable EE and SR increase linearly with the number of users (i.e. $K_{\rm Te}$), though the EE tends to be saturated, especially for PA-only and fixed PA-only.



In a comparison of the three system configurations across different user numbers, the proposed RIS+PA configuration consistently yields the highest EE and SR across all scenarios, maintaining a $12.7\%$ to $17.0\%$ EE (a $21.0\%$ to $38.3\%$ SR) advantage relative to PA-only (fixed PA-only). By the same token, it can also be observed that the PA-only configuration outperforms the fixed PA-only configuration uniformly for all the ($K_{\text{Tr}}$,$K_{\text{Te}}$) cases. Therefore, the above results demonstrate the effectiveness of the proposed GNN (Strategy I) for a practical beamforming solution and exhibit good reliability under any of the proposed three strategies.

\begin{table}[t]
\footnotesize
\belowrulesep=-0.3pt
\aboverulesep=-0.3pt
\centering
\caption{Performance evaluation of SR (bit/s/Hz) and EE (bit/J/Hz) of the proposed GNN for $K_{\rm Tr}\in\{2,4,6\}$, and $K_{\rm Te}= K_{\rm Tr}\pm 1$.}
\label{tab:generalization}
\renewcommand{\arraystretch}{1.15}
\setlength{\tabcolsep}{5pt}
\begin{tabular}{cc||cc|cc|cc}
\toprule
\multirow{2}{*}{\textbf{$K_{\rm Tr}$}} & \multirow{2}{*}{\textbf{$K_{\rm Te}$}} &
\multicolumn{2}{c|}{\textbf{RIS+PA}} &
\multicolumn{2}{c|}{\textbf{PA-only}} &
\multicolumn{2}{c}{\textbf{Fixed PA-only}} \\
\cmidrule(lr){3-4}\cmidrule(lr){5-6}\cmidrule(lr){7-8}
& & EE & SR & EE & SR & EE & SR \\
\midrule
\multirow{3}{*}{2} & 1 & 2.66 & 19.16 & 2.36 & 17.46 & 2.10 & 15.68 \\
& 2 & 4.98 & 35.87 & 4.39 & 32.41 & 3.84 & 29.54 \\
& 3 & 6.94 & 50.95 & 6.07 & 46.25 & 5.28 & 41.74 \\
 \hline
\multirow{3}{*}{4} & 3 & 6.92 & 50.88 & 6.10 & 46.25 & 5.31 & 41.81 \\
& 4 & 8.73 & 64.86 & 7.68 & 59.03 & 6.64 & 52.96 \\
& 5 & 10.26 & 77.50 & 8.99 & 70.10 & 7.69 & 62.71 \\
 \hline
\multirow{3}{*}{6} & 5 & 10.21 & 77.49 & 8.96 & 69.50 & 7.69 & 62.65 \\
& 6 & 11.45 & 88.22 & 10.00 & 78.91 & 8.51 & 70.84 \\
& 7 & 12.18 & 95.90 & 10.54 & 85.39 & 8.81 & 75.79 \\
\bottomrule
\end{tabular}
\begin{tablenotes}
\footnotesize
\item $K_{\rm {Tr}}$/$K_{\rm {Te}}$: Value of  $K$ in the training/test set.
\end{tablenotes}
\end{table}

\subsection{Impact of Key System Parameters}

This subsection discusses the impact of key system parameters, including the number of waveguides, the number of PAs, the length of waveguides, and the number of reflecting elements.

\subsubsection{Number of waveguides}

\begin{figure}[t]
    \includegraphics[width=0.50\textwidth]{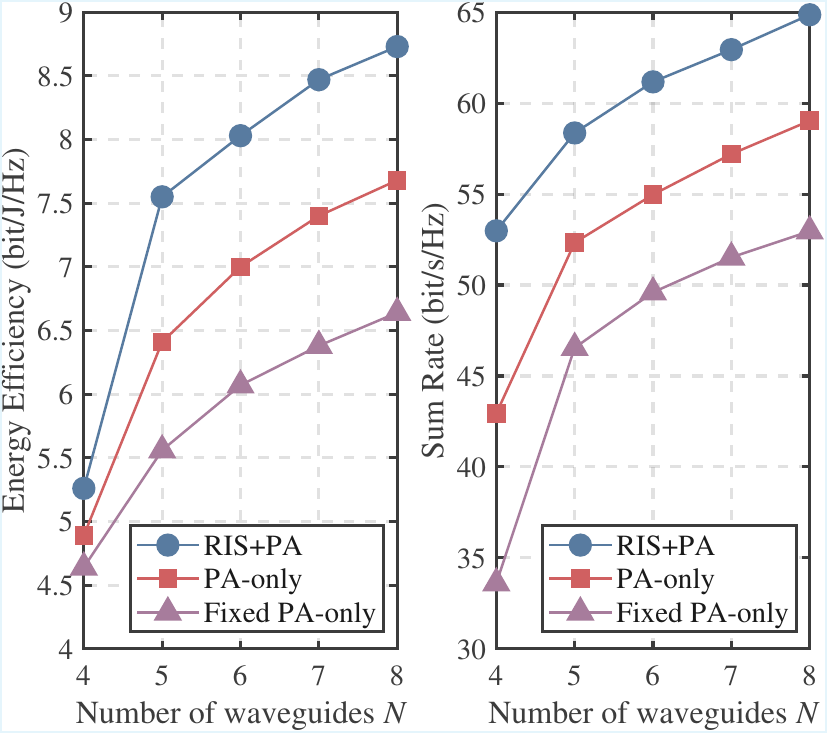}
    \caption{Impact of number of waveguides on EE and SR, with $L=32$, $M=8$ and $K=4$.}
    \label{waveguide_impact}
\end{figure}

Figure \ref{waveguide_impact} illustrates  the impact of the number of waveguides $N$ on EE and SR. As $N$ increases, all three system configurations exhibit improved EE and SR due to the increased availability of spatial degrees of freedom and reduced average path loss. Among them, RIS+PA still outperforms the other two configurations. Increasing $N$ is particularly beneficial for enhancing EE. Specifically, when $N$ increases form $4$ to $8$, RIS+PA achieves a $66\%$ EE improvement for RIS+PA, with the largest gains occurring at small $N$ scales. 

\subsubsection{Number of PAs}
\begin{figure}[t]
    \includegraphics[width=0.50\textwidth]{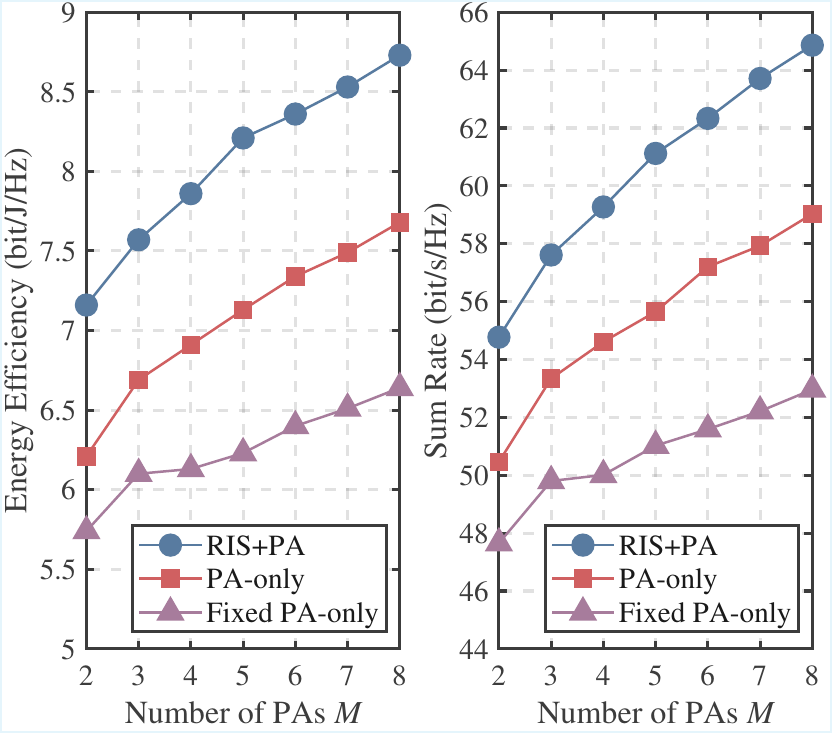}
    \caption{Impact of number of PAs on EE and SR, with $L=32$, $N=8$ and $K=4$.}
    \label{antenna_impact}
\end{figure}

Figure \ref{antenna_impact} illustrates  the impact of number of PAs $M$ on EE and SR. Similarly, EE and SR improve as $M$ increases, and RIS+PA achieves superior performance. Moreover, the performance gains of RIS+PA and PA-only over fixed PA-only are significantly enlarged, validating the effectiveness of optimizing PA placement. Notably, movable PAs  provide an additional degree of freedom compared to conventional fixed-position antenna systems, and a larger number of PAs can further exploit this advantage.


\subsubsection{Length of waveguide}

\begin{figure}[t]
    \includegraphics[width=0.50\textwidth]{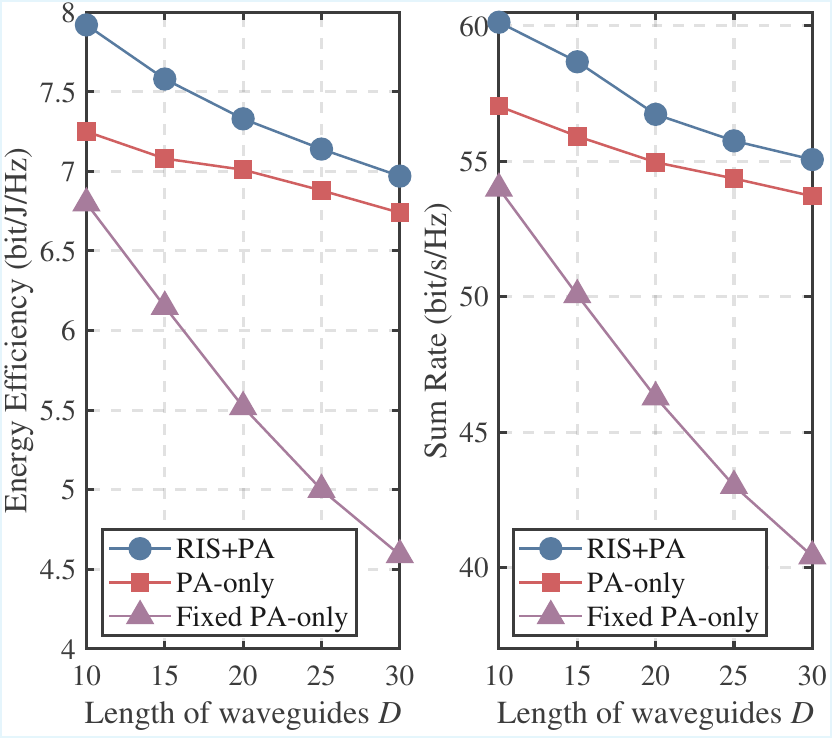}
    \caption{Impact of length of waveguide on performance with $N=8$, $L=32$, $M=8$ and $K=4$.}
    \label{area_distance}
\end{figure}

Figure \ref{area_distance} illustrates the impact of the length of waveguides $D$ (also the length of the waveguide deployment region along the $x$-axis) on EE and SR. A larger $D$ indicates that users are more sparsely distributed, which in turn increases the average path loss. Therefore, as $D$ increases, both EE and SR of all three configurations degrade despite the same performance rank. Two noticeable observations are: (O1) the performance gap between RIS+PA and PA-only gets smaller with $D$, and (O2) that between PA-only and fixed PA-only increases significantly with $D$. The reason for (O1) is that RIS enables a cascaded communication link at the cost of multiplicative fading for a larger waveguide deployment region. Moreover, (O2) indicates that movable PAs can compensate for the performance loss resulting from severer path loss.

\subsubsection{Number of reflecting elements}

\begin{figure}[t]
    \includegraphics[width=0.50\textwidth]{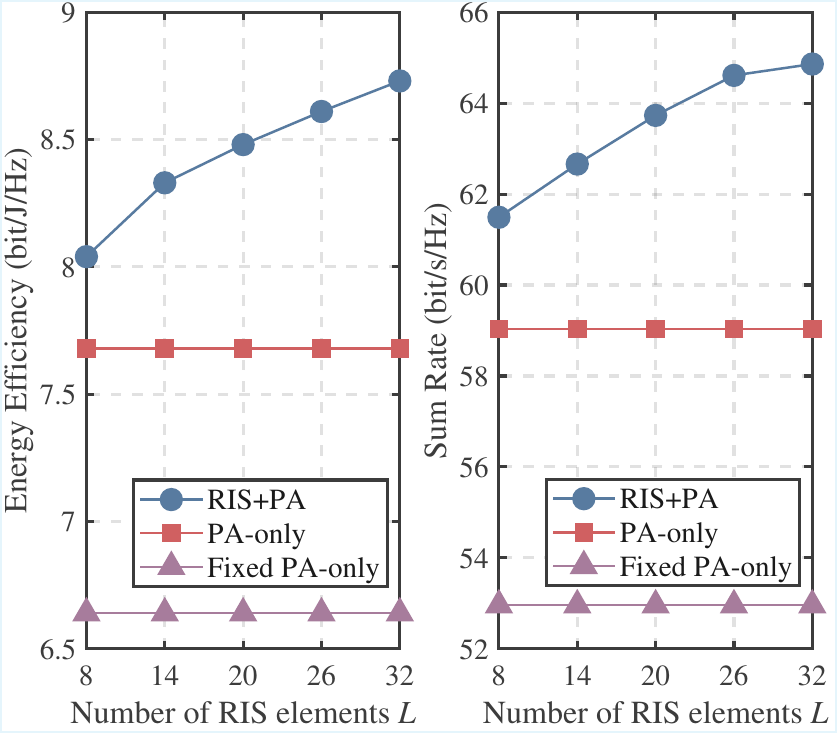}
    \caption{Impact of number of reflecting elements on performance with $N=8$, $M=8$ and $K=4$.}
    \label{ris_size}
\end{figure}

Figure \ref{ris_size} illustrates the impact of the number of RIS reflecting elements $L$ on EE and SR. Varying $L$ only affects RIS+PA, as PA-only and Fixed PA-only do not involve RIS. In general, increasing $L$ improves the system performance of RIS+PA, but the performance gain gradually saturates. Apparently, quadrupling the RIS size yields less than a $9\%$ EE gain, suggesting that beyond $L=26$, additional reflecting elements contribute marginal benefits for this case. 


\subsection{Inference time}

\begin{table}[t]
\footnotesize
\belowrulesep=-0.3pt
\aboverulesep=-0.3pt
\centering
\caption{Inference times (in milliseconds) of the proposed three-stage GNN (Strategy I), MLP, and CVX for two cases of $(N,L,M,K)$.}
\label{inference_time}
\renewcommand{\arraystretch}{1.15}
\setlength{\tabcolsep}{5pt}
\begin{tabular}{c||cc|cc|cc}
\toprule
\multirow{2}{*}{$(N,L,M,K)$}  & 
\multicolumn{2}{c|}{\textbf{MLP}} &
\multicolumn{2}{c|}{\textbf{GNN}} &
\multicolumn{2}{c}{\textbf{CVX}} \\
\cmidrule(lr){2-3} \cmidrule(lr){4-5} \cmidrule(lr){6-7}
 & EE & SR & EE & SR & EE & SR \\
\midrule
$(2,8,2,2)$ & 0.02  & 0.02 & 0.33  & 0.33 & $3.01\times10^3$ & $3.09\times10^3$ \\
$(8,32,8,6)$ & 0.02  & 0.02  & 1.78  & 1.77  & $5.16\times10^3$ & $1.41\times10^4$ \\
\bottomrule
\end{tabular}
\end{table}

Table \ref{inference_time} compares inference time across different methods under two scenarios.  As observed, learning-based approaches achieve a remarkable speedup of three orders of magnitude over CVX, thereby enabling real-time deployment. As the problem size increases, all the inference times of learning-based methods grow slightly. In contrast, CVX experiences a tremendous growth in inference time (thus not real-time applicable). While the GNN incurs a slightly higher computational cost than the MLP due to its message-passing mechanisms, its inference time remains well below 2 milliseconds even for large-scale scenarios. The advantages of GNN over MLP  lie in better system performance and  generalization capability for unseen problem sizes without retraining, as discussed above, thus exhibiting high flexibility in transmission design.




\section{Conclusion}
For the multi-user downlink transmission of a RIS-assisted multi-waveguide PASS system (cf. Fig. \ref{sys}), we have presented an unsupervised three-stage GNN, constituted by three sub-GNNs in cascade (PAGNN for generating PA positions, RISGNN for generating the phase shift matrix, and BeamGNN for generating the beamforming vectors) as shown in Fig. \ref{structure}. Additionally, we proposed three implementation strategies for integrating the GNN with conventional optimization algorithms that offer different trade-offs between computational complexity and solution optimality. Extensive numerical results have demonstrated the three-stage GNN's superior performance as well as its unique attributes: high design flexibility, viable generalization, and good performance reliability. Notably, within millisecond-level inference time, it can yield a high-quality solution to either of EE and SR maximization problems (cf. \eqref{p1:2a1} and \eqref{p1:2a2}), thus exhibiting its real-time applicability. As a final remark, to the best of our knowledge, no existing effective transmission  designs have been reported in the open literature for the PASS system under the challenging scenario considered in our work, so the proposed GNN is the first practical solution for future wireless communications.

\end{document}